\documentclass[a4paper]{easychair}

\usepackage{doc}
\usepackage{makeidx}
\usepackage{amssymb}
\usepackage[figuresright]{rotating}
\usepackage{url,lineno}

\urldef{\mailsa}\path|{ivanherreros@gmail.com}| 

\begin{document}

\title{\bf  Cerebellar-Inspired Learning Rule for Gain Adaptation of Feedback Controllers}
\titlerunning{ }
\author{Ivan Herreros\inst{1} \and Xerxes D. Arsiwalla\inst{1}  \and  Cosimo Della Santina\inst{2} \and Jordi-Ysard Puigbo\inst{1} \and Antonio Bicchi\inst{2,3} \and Paul   Verschure\inst{1,4} }
\authorrunning{ }
\institute{SPECS Lab, University Pompeu Fabra \\
		Carrer de Roc Boronat, 138, 08018 Barcelona, Spain
\and
Research Center ``En\-ri\-co Pi\-ag\-gio'', U\-ni\-ver\-si\-ty of Pisa \\
		Largo Lucio Lazzarino 1, 56126 Pisa, Italy
\and
Department of Advanced Robotics, Istituto Italiano di Tecnologia \\
via Morego, 30, 16163 Genoa, Italy
\and
Instituci\'{o} Catalana de Recerca i Estudis Avan\c{c}ats \\
Passeig de Lluís Companys, 23, 08010 Barcelona, Spain \\
\mailsa\\ 
}
\date{}

\maketitle



%

\begin{abstract}
How does our nervous system successfully acquire feedback control strategies in spite of a wide spectrum of response dynamics from different musculo-skeletal systems? The cerebellum is a crucial brain structure in enabling precise motor control in animals. Recent advances suggest that synaptic plasticity of cerebellar Purkinje cells involves molecular mechanisms that mimic the dynamics of the efferent motor system that they control allowing them to match the timing of their learning rule to behavior. Counter-factual predictive control (CFPC) is a cerebellum-based feed-forward control scheme that exploits that principle for acquiring anticipatory actions. CFPC extends the classical Widrow-Hoff/Least Mean Squares by inserting a forward model of the downstream closed-loop system in its learning rule. Here we apply that same insight to the problem of learning the gains of a feedback controller. To that end, we frame a model-reference adaptive control (MRAC) problem and derive an adaptive control scheme treating the gains of a feedback controller as if they were the weights of an adaptive linear unit. Our results demonstrate that rather than being exclusively confined to cerebellar learning, the approach of controlling plasticity with a forward model of the subsystem controlled, an approach that we term as Model-enhanced least mean squares (ME-LMS), can provide a solution to wide set of adaptive control problems.
\end{abstract}


\section{Introduction}
The cerebellum is arguably the brain structure whose study has had a deeper impact on the robotics and control communities. The seminal theory of cerebellar function by Marr \cite{Marr1969} and Albus \cite{albus1971theory} was translated by the latter into the cerebellar model articulation controller (CMAC) in the early seventies \cite{Albus1975}, which up until today is used both in research and applications. A decade later, Fujita \cite{Fujita1982} advanced the adaptive filter theory of cerebellar function based on the work by Widrow et al. \cite{widrow1960adaptive}. Later, in the late eighties, Kawato formulated the influential feedback error learning (FEL) model of cerebellar function \cite{Kawato1987}, in which the cerebellum, implemented as an adaptive filter, learned from, and supplemented, a feedback controller. Unlike CMAC, FEL had a strong impact within the neuroscientific community as a theory of biological motor control \cite{Wolpert1998a}. Within the robotics and control communities, FEL has been studied in terms of performance and convergence properties \cite{nakanishi2004feedback}. Later, the adaptive filter theory of cerebellar function was revived by Porrill and colleagues \cite{Dean2010}, proposing alternatives to FEL that have been applied to the control of bio-mimetic actuators, like pneumatic or elastomer muscles \cite{lenz2009cerebellar},  \cite{wilson2016cerebellar}.  

More recently, the counterfactual predictive control (CFPC) scheme was proposed in \cite{herreros2016forward}, motivated from neuro-anatomy and physiology of eye-blink conditioning, a behavior dependent on the cerebellum. CFPC includes a reactive controller, which is an output-error feedback controller that models brain stem or spinal reflexes actuating on peripheral muscles, and a feed-forward adaptive component that models the cerebellum and learns to associate its own inputs with the errors that drive the reactive controller. CFPC proposes that the learning of adaptive terms in the linear filter should depend on the coincidence of an error signal with the output of a forward model implemented at the synaptic level, reproducing the dynamics of the downstream reactive closed-loop system. We refer to that learning rule as a model-enhanced least-mean squares (ME-LMS) rule to differentiate with the basic least-mean squares (LMS) rule that is implemented in previous models of the cerebellum, such as CMAC and FEL. In agreement with the theoretical insight of CFPC, recent physiological evidence in \cite{suvrathan2016timing} shown that the timing of the plasticity rule of Purkinje cells is matched to behavioral function. That suggests that Purkinje cells, the main cells implied in learning at the level of the cerebellum, have  plasticity rules that reflect the sensorimotor latencies and dynamics of the plants they control.
%

However, in the context of CFPC, the ME-LMS rule was derived as a batch gradient-descent rule for solving a feed-forward control task in discrete time. In that sense, it can be interpreted as providing a solution to a iterative-learning control scheme, an input design technique for learning to optimize the execution of a repetitive task \cite{bristow2006survey}. Hence, it remained open the question as to whether a similar learning rule could support the acquisition of well-tuned feedback gains. That is, whether ME-LMS could be applied in an adaptive \emph{feedback} control problem. Here we answer that using the model reference adaptive control (MRAC) frame \cite{aastrom1983theory}. In that frame, we first show that the biologically-inspired ME-LMS algorithm can be derived from first principles. More concretely, we show that the ME-LMS rule emerges from deriving the stochastic gradient descent rule for the general problem of updating the gains of linear proportional feedback controllers actuating on a LTI system. Finally we test in simulation the effectiveness of the proposed cerebellum-inspired architecture in controlling a damped-spring mass system, a non-minimum phase plant and, finally, closing the loop with the biology, a biologically-based model of a human limb.

\section{Derivation of the ME-LMS Learning Rule}

\begin{figure}[h]
	\centering
	\begin{minipage}[h]{0.33\linewidth} 
		\centering
		\includegraphics[width=0.99\linewidth]{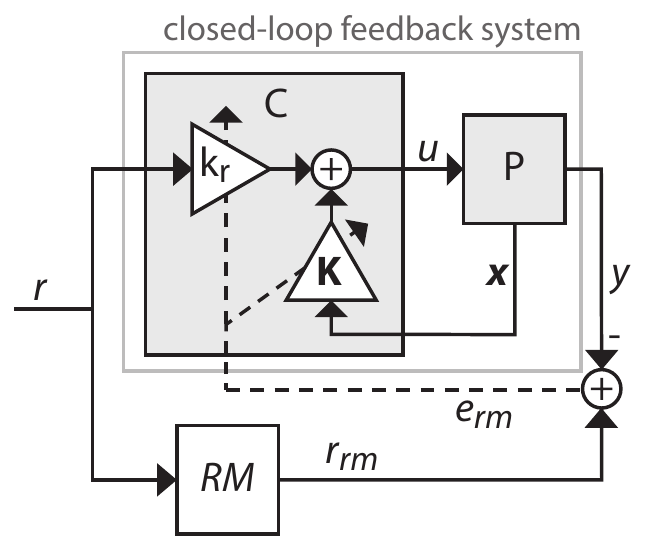}
	\end{minipage}%
	\begin{minipage}[h]{0.33\linewidth} 
		\centering
		\includegraphics[width=0.99\linewidth]{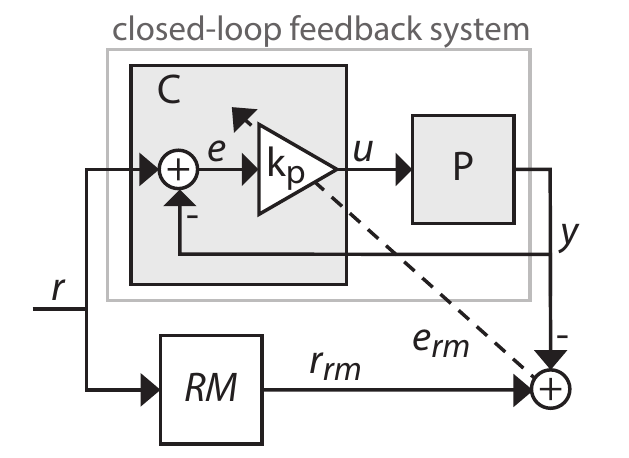} 
	\end{minipage}
	\caption{Adaptive architecture for the full state feedback (\emph{left}) and output-error proportional (P, \emph{right}) control cases. Abbreviations: $\mathrm{C}$, feedback controller, $\mathrm{P}$, plant; $\mathrm{RM}$, reference model; $r$, reference signal; $r$, reference signal outputted by the reference model; $y$, plant's output; $e$, output error; $e_{rm}$, output error relative to the reference model; $\mathbf{x}$, state of the plant; $u$, control signal; $k_r$, reference gain; $\mathbf{K}$, state feedback gains; and $k_p$, proportional error-feedback gain.}\label{fig:generic_architectures}
\end{figure}

In the next we derive a learning rule for learning the controller gains for both state and output-error feedback controllers. The generic architectures for a full-state-feedback and a proportional (P) controller are shown in Fig. \ref{fig:generic_architectures}. To define the model-reference adaptive control (MRAC) problem, we set a reference model whose output we denote by $r_{rm}$. The error in following the output of the reference model, $e_{rm}=r_{rm}-y$, drives adaptation of the feedback gains. But note that in the proportional error feedback controller, $e=r-y$ is the signal feeding the error feedback gain.


\subsection{ME-LMS for State-Feedback}

For the derivation purposes we  assume that the adaptive control strategy is applied to a generic LTI dynamical system
\begin{equation}
\begin{array}{rcl}
{\dot{\mathbf{x}}}&=&  \mathbf{Ax}+\mathbf{B}u\\
y               &=&  \mathbf{Cx}
\end{array}
\label{eq:LDS}
\end{equation}
where $\mathbf{A}\in\mathbb{R}^{N\times N}$, $\mathbf{B}\in\mathbb{R}^{N\times 1}$ and $\mathbf{C}\in\mathbb{R}^{1\times N}$ are the usual dynamics, input and output matrices, respectively; $\mathbf{x}\in\mathbb{R}^N$ is the state vector; and $u$ and $y$, both scalars, are the input and output signals. 

The control signal will be generated according to the following state-feedback law
\begin{equation}
\begin{array}{rcl}
u               &=& \mathbf{Kx}+k_rr
\end{array}
\end{equation}
where $r$ is the reference signal, $\mathbf{K}\in\mathbb{R}^{1\times N}$ is the (row) vector of state feedback gains and $k_r$ the reference gain. Both $\mathbf{K}$ and $k_r$ are here time-dependent and will be updated by the learning rule controlling adaptation.

Substituting the control law within the dynamics equation, we obtain the closed-loop system description
\begin{eqnarray}
\dot{\mathbf{x}}&=&  (\mathbf{A}+\mathbf{BK})\mathbf{x}+\mathbf{B}k_rr\\
y               &=&  \mathbf{Cx}
\end{eqnarray}
We set the L-2 norm of the error as the cost function to minimize \[J=\frac{1}{2}e_{rm}^2\label{2_cost}\]

For convenience, we write now the control law as $u=\tilde{\mathbf{K}}^\mathsf{T}\tilde{\mathbf{x}}$ with $\tilde{\mathbf{K}}\in\mathbb{R}^{N+1} \equiv [k_1, \dots, k_N, k_r]^\mathsf{T}$ and $\tilde{\mathbf{x}}\in\mathbb{R}^{N+1} \equiv [x_1,\dots,x_N,r]^\mathsf{T}$. 
To derive the gradient descent algorithm for adjusting the vector of gains, $\tilde{\mathbf{K}}$, we need the gradient of $J$ with respect to $\tilde{\mathbf{K}}$:

\begin{equation}
\nabla_{\tilde{\mathbf{K}}} J = \frac{\partial e_{rm}}{\partial \tilde{\mathbf{K}}} e_{rm} = 
-\frac{\partial y}{\partial \tilde{\mathbf{K}}} e_{rm}
\end{equation}

Now we will consider each of the gains individually, treating separately the state and the reference gains. Let $k_i$ denote the feedback gain associated with the i-th state variable. We have that 

\begin{equation}
\frac{\partial y}{\partial k_i} = \mathbf{C}\frac{\partial \mathbf{x}}{\partial k_i}
\end{equation}

We compute the partial derivative of the state vector $\mathbf{x}$ with respect to  $k_i$ applying the partial derivative to the differential equation that governs the closed-loop dynamics:

\begin{equation}
\frac{\partial}{\partial k_i} \dot{\mathbf{x}}= 
\frac{\partial}{\partial k_i} \left((\mathbf{A}-\mathbf{BK})\mathbf{x}+\mathbf{B}k_rr\right)
\end{equation}

Using the substitution $\mathbf{z}_i \equiv \frac{\partial \mathbf{x}}{\partial k_i}$ and applying the product rule in the derivation we obtain

\begin{equation}
\dot{\mathbf{z}}_i= (\mathbf{A}-\mathbf{BK})\mathbf{z}_i+\mathbf{B}x_i
\end{equation}

Introducing $h_i \equiv \mathbf{Cz}_i$, we get \[\frac{\partial J}{\partial k_i}=h_i e_{rm}\] Note that this has solved the problem of obtaining the partial derivative for all state feedback gains.

In the case of the reference gain, with $\mathbf{z}_r \equiv \frac{\partial \mathbf{x}}{\partial k_r}$, we obtain  

\begin{equation}
\dot{\mathbf{z}}_r= (\mathbf{A}-\mathbf{BK})\mathbf{z}_r+\mathbf{B}r
\end{equation}

And introducing $h_r \equiv \mathbf{Cz}_r$, \[\frac{\partial J}{\partial k_r}=h_r e\]

We will refer to the quantities $h_i$ and $h_r$ as eligibility traces. We can write the vector of eligibility traces as follows: $\tilde{\mathbf{h}} = [h_1, \dots, h_N, h_{r}]^\mathsf{T}$.

With this we can solve for the gradient of the cost function as follows
\[\nabla_{\tilde{\mathbf{K}}} J = - \tilde{\mathbf{h}} e_{rm}\]
And consequently derive a learning rule for the gains that will follow a gradient descent:

\begin{equation}
\dot{\tilde{\mathbf{K}}} = \eta \tilde{\mathbf{h}} e_{rm}
\label{eq_MELMS}
\end{equation}

Note that this rule is similar to the classical Widrow-Hoff or least mean squares (LMS) rule. However, in the standard LMS, the rate of change is  obtained multiplying the error with the input signals of the filter ($\dot{\tilde{\mathbf{K}}} = \eta \tilde{\mathbf{x}} e_{rm}$) whereas in the rule we have derived the error multiplies the quantities in $\tilde{\mathbf{h}}$, which are obtained after passing the input signals through a forward model of the controlled system. For this reason, we refer to the learning rule in equation \ref{eq_MELMS} as model-enhanced least mean squares (ME-LMS). Moreover, $\tilde{h}_i$ is the eligibility trace associated to the input $\tilde{x}_i$ because, at the time that a particular error signal comes, it codes how much $\tilde{x}_i$ could have contributed to canceling that error.

\subsection{ME-LMS for Output Error Proportional Control}

The control law of an output-error proportional controller is $u=k_p e=k_pr-k_p\mathbf{Cx}$. Following a derivation analogous to the previous one, we obtain the following expression for computing eligibility trace of the proportional gain ($h_p$).
\begin{eqnarray}
\dot{\mathbf{z}}_p&=&  (\mathbf{A}-\mathbf{BC}k_p)\mathbf{z}_p+\mathbf{B}r\\
h_p               &=&  \mathbf{Cz}_p
\end{eqnarray} 
hence, in this model, plasticity will be implemented by the rule $\dot{k}_p = \eta h_p e$.

\subsection{Model-Enhanced Least Mean-Squares vs. Least Mean Squares Rule}

\begin{figure}
	 \centering
	\includegraphics[width = 0.67\columnwidth]{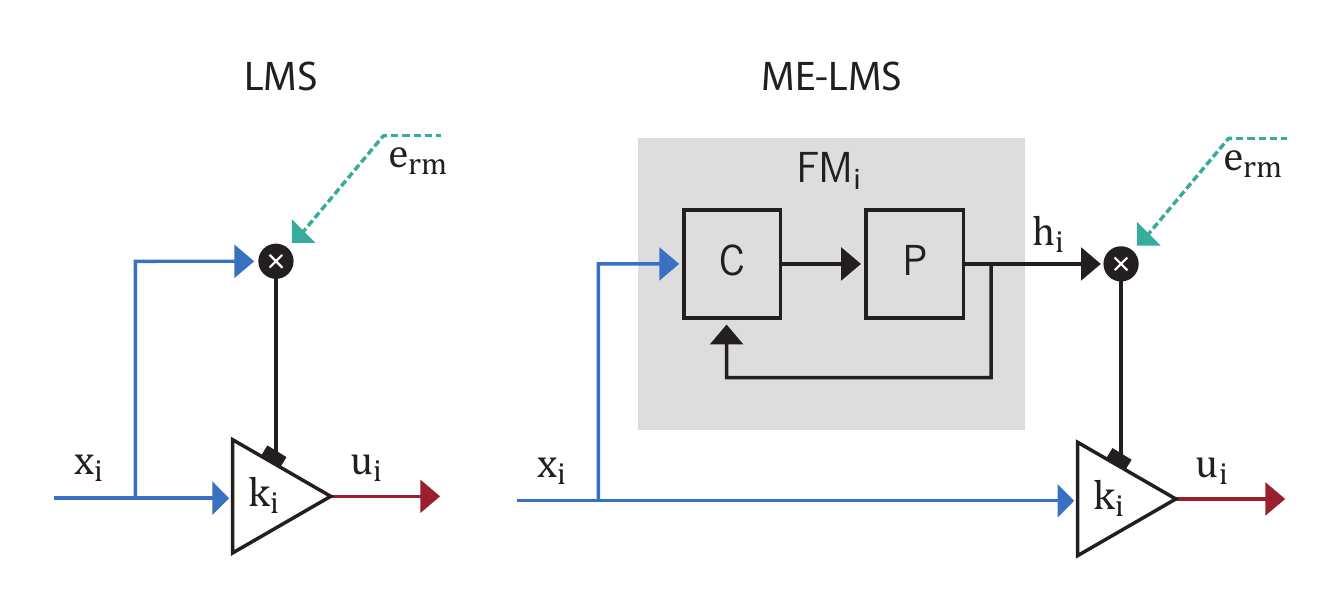}
	\caption{Schematic of the implementation of the LMS (left) and the ME-LMS (right) rule at the level of a single adaptive weight. Note that since $k_i$ is a gain of the controller $C$, this scheme is implicitly recursive: as the $k_i$ gain changes (together with the other adaptive gains of the controller) the forward that it utilizes to drive plasticity, changes as well.}\label{fig:lms_melms}
\end{figure}

The differences between LMS and ME-LMS care summarized as follows: In LMS, the change of the adaptive gain $k_i$ is determined based on the temporal coincidence of a global error signal $e_{rm}$ and the local input $x_i$ (Fig. \ref{fig:lms_melms} \textit{left}). In ME-LMS (Fig. \ref{fig:lms_melms} \textit{right}) the change in the gain is controlled by the output of a gain-specific forward model $FM_i$, whose output $h_i$ facilitates an eligibility trace for  $k_i$. The term \emph{eligibility trace} implies that $h_i$ marks how much the input $x_i$ could have contributed to decrease the current error. In that sense, it is a \emph{trace} as long as to be generated $h_i$ takes into account the history of $x_i$ with a time-span implicitly determined by the dynamics of the forward model.

\section{Applying ME-LMS to a Linear Damped Spring-Mass System}

We evaluate here the performance of the proposed algorithm in controlling a standard damped-spring mass system: $m\ddot{q} + c\dot{q} + kq = u$, with $m=1$Kg, $c=0.5\,\frac{\text{Ns}}{\text{m}}$, $k=0.5 \,\frac{\text{N}}{\text{m}}$.

\subsection{Output Error P-Control}
\begin{figure}[h]
	\centering
	\begin{minipage}[h]{0.33\linewidth} 
		\centering
		\includegraphics[width=0.99\linewidth]{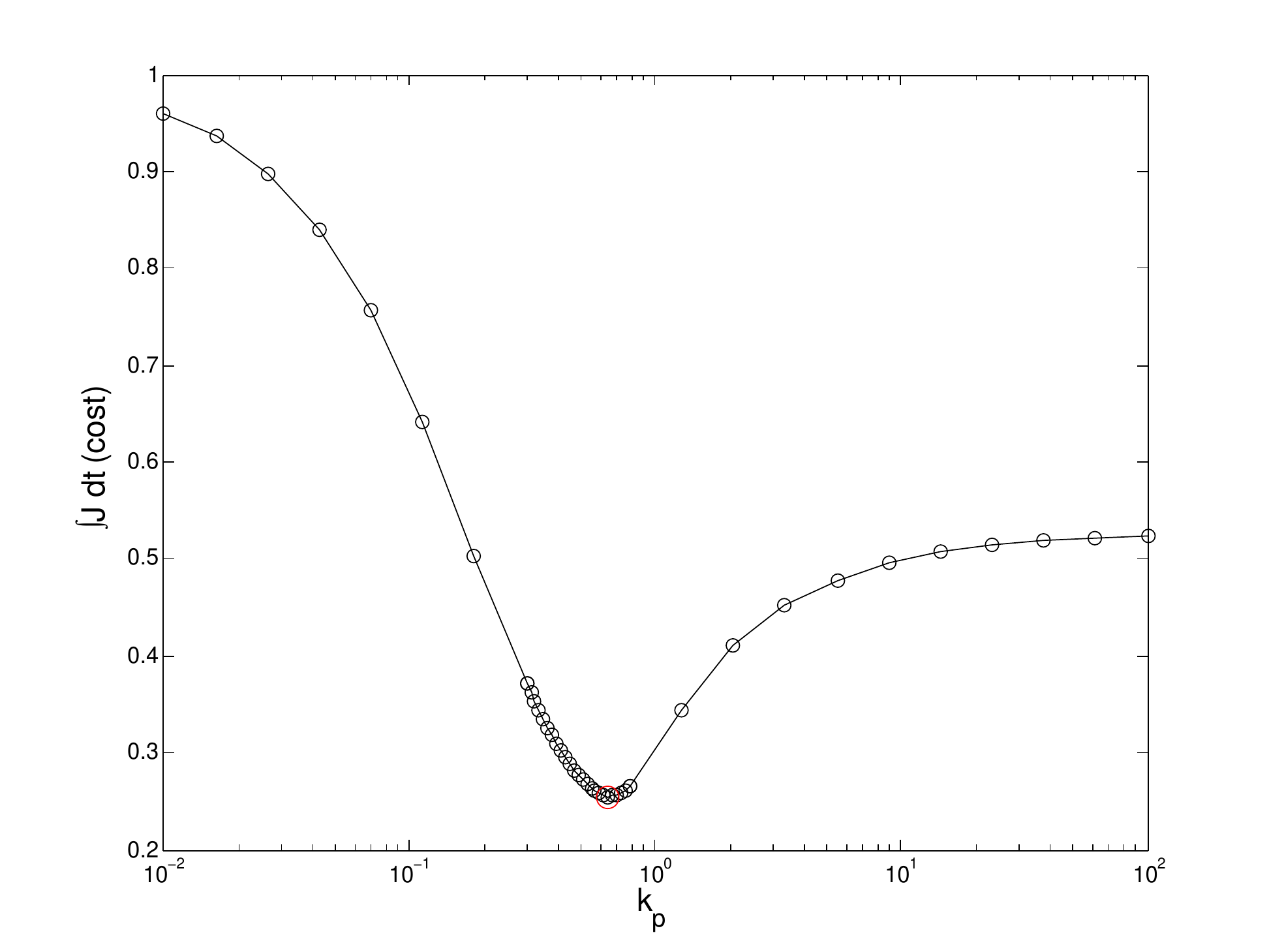}
		\includegraphics[width=0.99\linewidth]{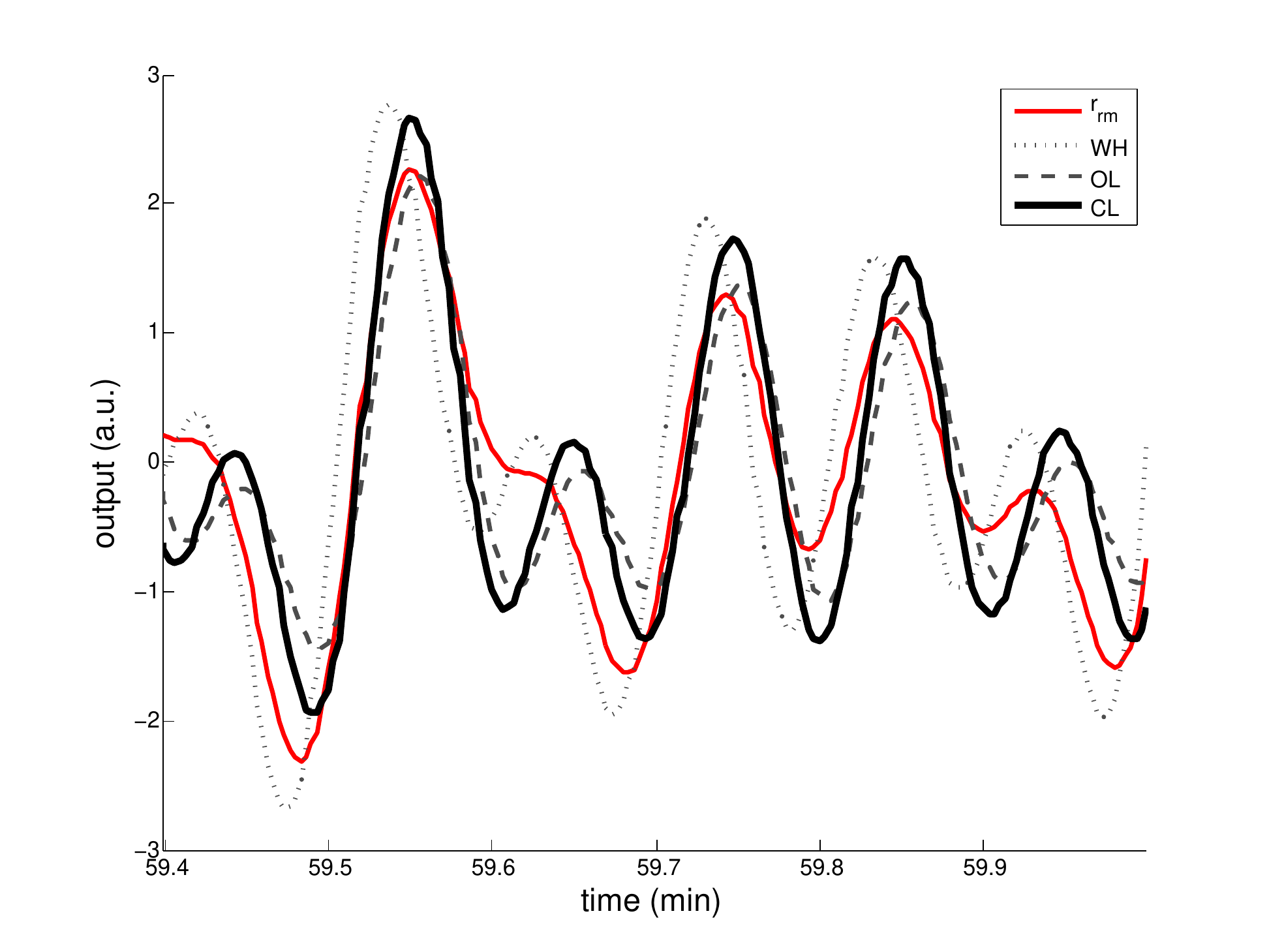} 
	\end{minipage}%
	\begin{minipage}[h]{0.33\linewidth} 
		\centering
		\includegraphics[width=0.99\linewidth]{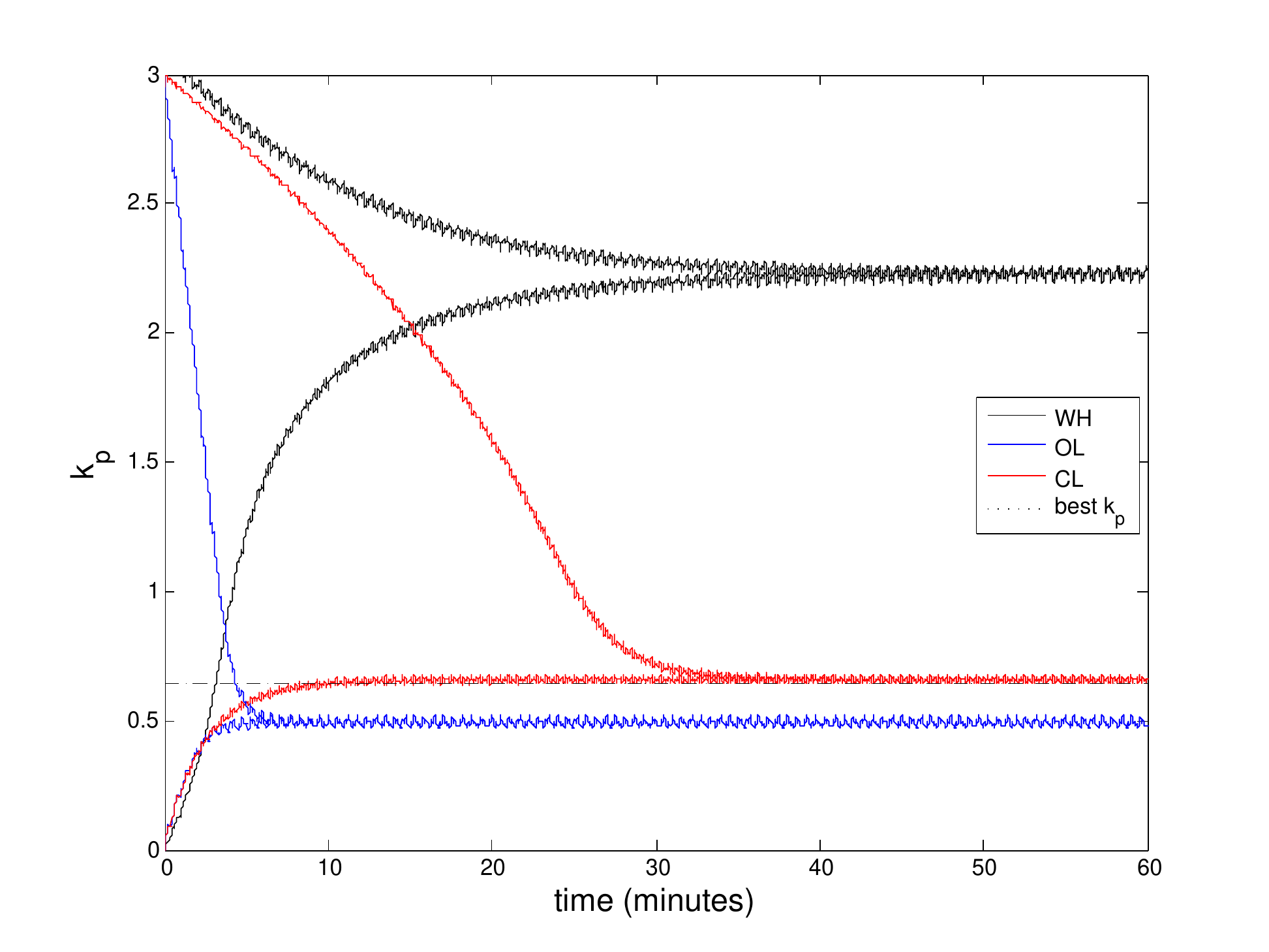} 
		\includegraphics[width=0.99\linewidth]{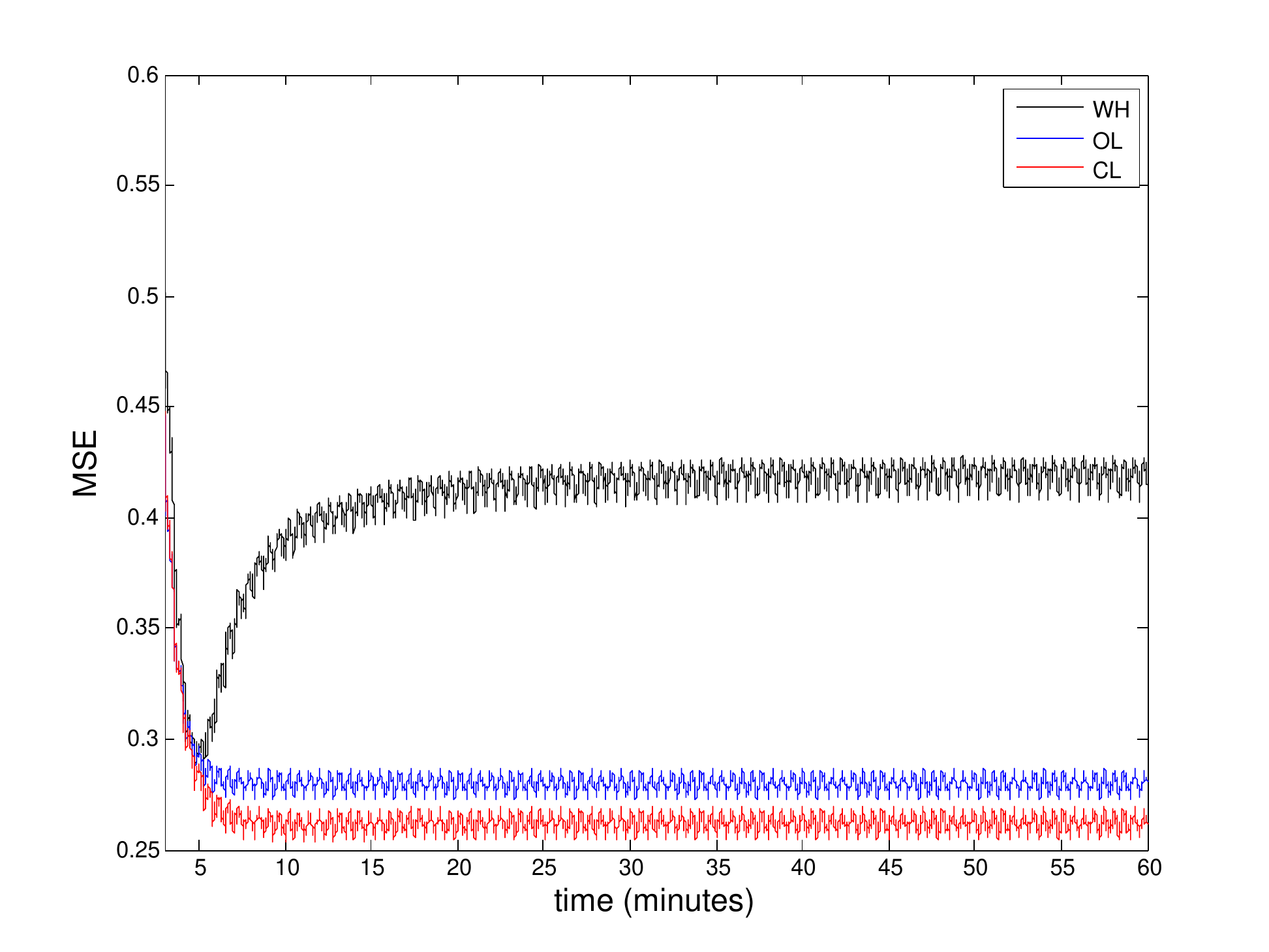} 
	\end{minipage}
	
	\caption{Damped-spring mass system with P-control. \textit{Above left}. Cost as a function of $k_p$. \textit{Above right}. Convergence of $k_p$ for three different models. \textit{Below left}. Output trajectories and reference signals. \textit{Below right}. Evolution of the error in mean-square error (MSE) units. \textit{In all panels}: WH: standard LMS rule; OL: LMS with eligibility trace derived from the open-loop system; CL: ME-LMS with eligibility trace derived from the closed-loop system.}
	\label{fig:P-spring-mass}
\end{figure}

For this problem we use a reference model built as chain of two leaky integrators with identical relaxation time constants ($\tau=\sqrt{0.5} s$). The impulse response curve of this reference model corresponds to a normalized double-exponential convolution that peaks at $0.5 s$. Finally, we use as reference the following superposition two sinusoidal functions $r = \sin(5t/7)+\sin(5t/11)$. We first examine the cost function as a function of the feedback parameter, $k_p$ varying it logarithmically from  $0.01$ to $100.0$. Within that range, the cost function is convex and has minimum in the near $k_p=0.815$ (Fig \ref{fig:P-spring-mass} \textit{above-left}). At this point we check whether the ME-LMS converges to the optimum value for $k_p$. For comparison we also run the test with using the standard LMS rule, and a heuristically motivated alternative algorithm wherein we use a model of the open-loop system to generate the eligibility trace. We test two starting values each one at a different side of the minimum and we observe that in both cases the ME-LMS converges around the optimal $k_p$ (Fig \ref{fig:P-spring-mass} \textit{above-right}) while the alternative algorithms convergence to different equilibrium points which are non-optimal in cost terms. The difference in performance can also be appreciated by seeing how the different algorithms track $r_{rm}$ at the end of the training period (1$h$ of simulated time) (Fig \ref{fig:P-spring-mass} \textit{below-left}). Indeed, only the ME-LMS algorithm is in-phase with $r_{rm}$. Finally, in cost terms, only ME-LMS converges rapidly to the minimum value (Fig \ref{fig:P-spring-mass} \textit{above-right}). 

In summary, this result shows that even for the simplest feedback learning scenario, a LMS-like learning rule  converges to the optimal gain only if it uses the eligibility trace generated by a forward model that reproduces the dynamics of the closed-loop system.

\section{Applying ME-LMS to a Non-Minimum Phase Plant}
\subsection{Full State-Feedback Control}

In this section we apply ME-LMS to a non-minimum phase system, which is a system with zeros in the right-hand side of the complex plane. Acting to reduce an error in such a system requires foresight in that before minimizing an error one has to steer the plant in the direction that increases it. That property of the system, namely that errors cannot be reduced instantaneously, disqualifies the use of the standard LMS algorithm for the problem of adaptively tuning feedback gains. On the contrary, ME-LMS, as it takes explicitly into account the dynamics of the controlled system to control plasticity, can in principle appropriately adjust the gains of a feedback controller even when it is applied to a non-minimum phase system.

As a non-minimum phase system, we use the following a linearized model of a balance system (e.g., a self-balancing robot):

\[
\mathbf{A}=\left[ 
\begin{array}{ccc}
0 & 0 & 1 \\
m^2 l^2 g / \mu &  -c J_t / \mu & -\gamma l m / \mu  \\ 
M_t m g l / \mu  & -c l m / \mu  &  -\gamma M_t / \mu \\
\end{array}
\right]
\]\[ 
\mathbf{B}=\left[ 
\begin{array}{c}
0 \\ J_t / \mu \\ l m / \mu 
\end{array}
\right], 
\mathbf{C}=\left[ 
\begin{array}{ccc}
0 &  1 & 0
\end{array}
\right]
\]
where $\mu = M_t J_t - m^2 l^2$. The values, chosen to mimic the ones of a custom made prototype, are $M_t=1.58 Kg$, $m=1.54 Kg$, $l=0.035$, $J_t=1.98\times10^{-3}$, 	$\gamma=0.01$ and $c=0.1$. As an added difficulty, the plant is unstable in open loop. To deal with that, we set boundary conditions to our simulation. That is, whenever the system reaches a threshold velocity of $0.5 m/s$ or and inclination in absolute value above $22.5$ degrees the simulation  re-starts and the system is brought back to the initial rest position. In that sense, the system is not fully autonomous but \emph{assisted} by an external agent that makes it regain the balanced position.

In practice, the control problem consisted in following a low amplitude and slow velocity reference signal constructed as a sum of sinusoidals $0.05(\sin(\pi t/70)+\sin(\pi t/110))$. We used the same reference model as in the previous section. 

For this system the problem of adjusting the velocity to a particular reference signal is under-constrained as there are two possible strategies: keeping the error in the linear velocity equal to zero while the angular position diverges or keeping that error equal to zero while maintaining the angular position stable. In order to bias the system towards the second solution we set the starting gains already tuned towards the right solution. However, that initial set of gains keep the robot balanced for less than $200 ms$. Hence, we can divide this particular problem in two stages: first stabilizing the plant, and next make the controlled system converge to the dynamics of the reference model. 

\begin{figure}[h]
  \centering
  \begin{minipage}[h]{0.33\linewidth} 
\centering
\includegraphics[width=0.99\linewidth]{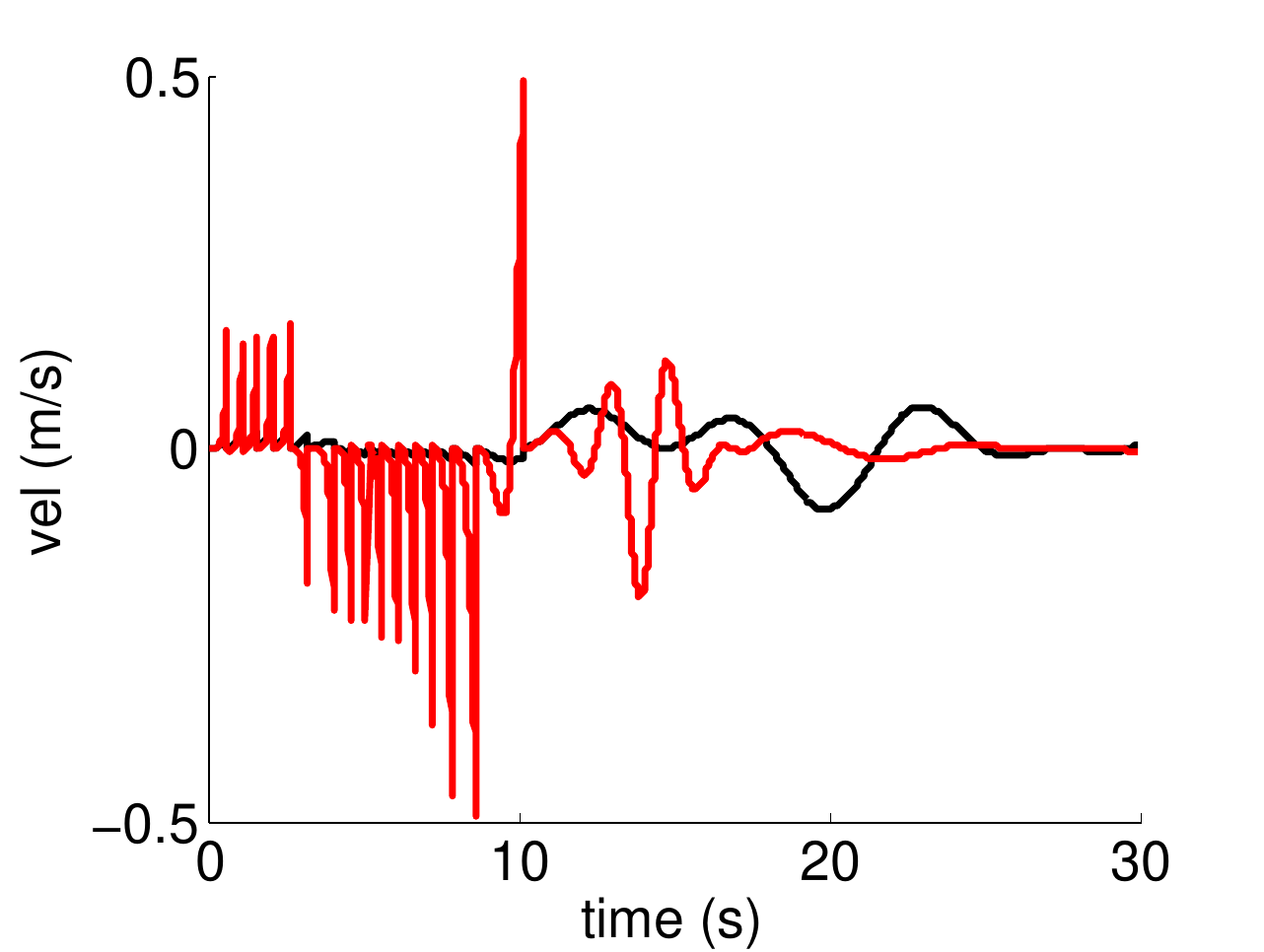}
\includegraphics[width=0.99\linewidth]{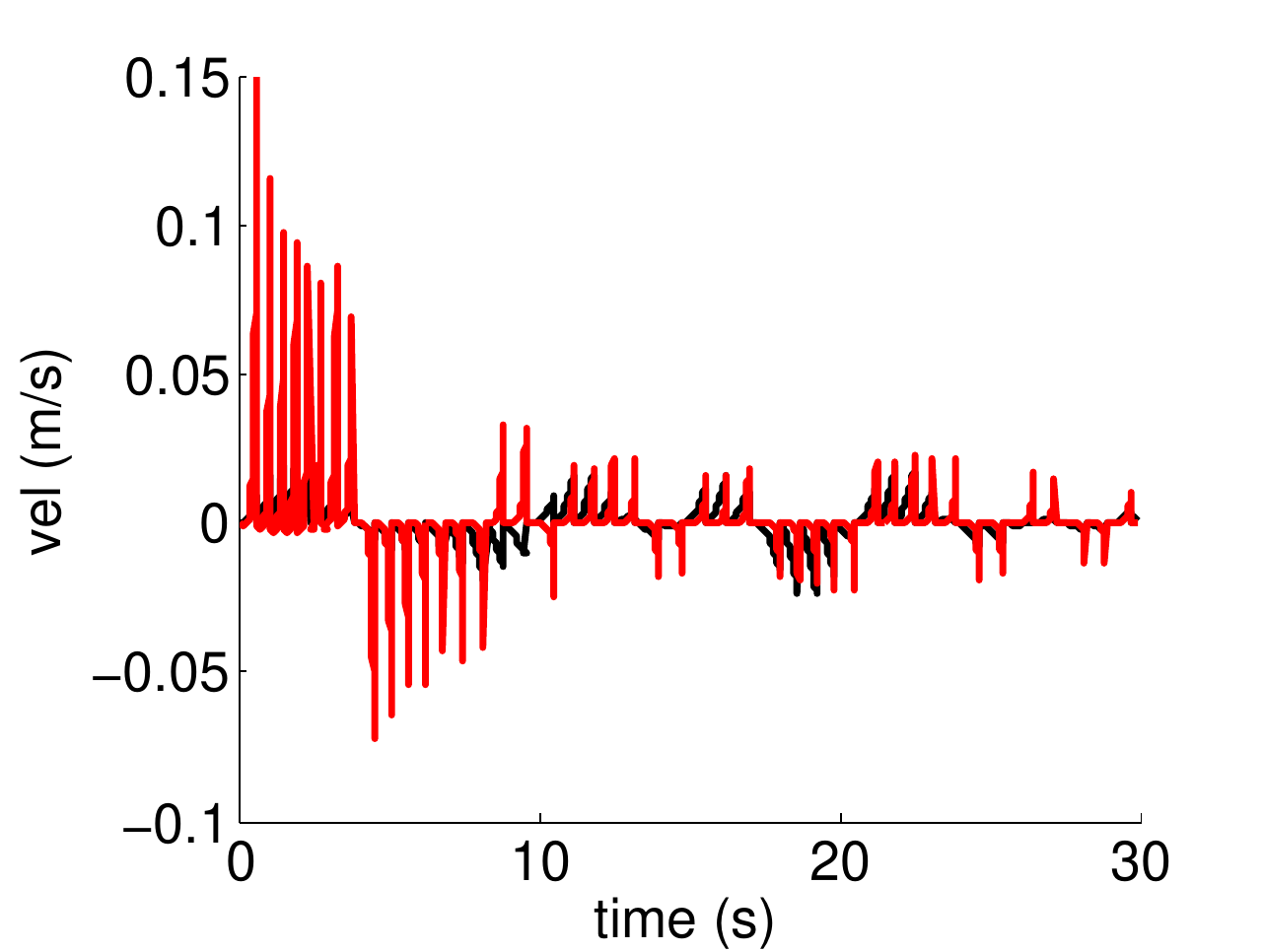}
\includegraphics[width=0.99\linewidth]{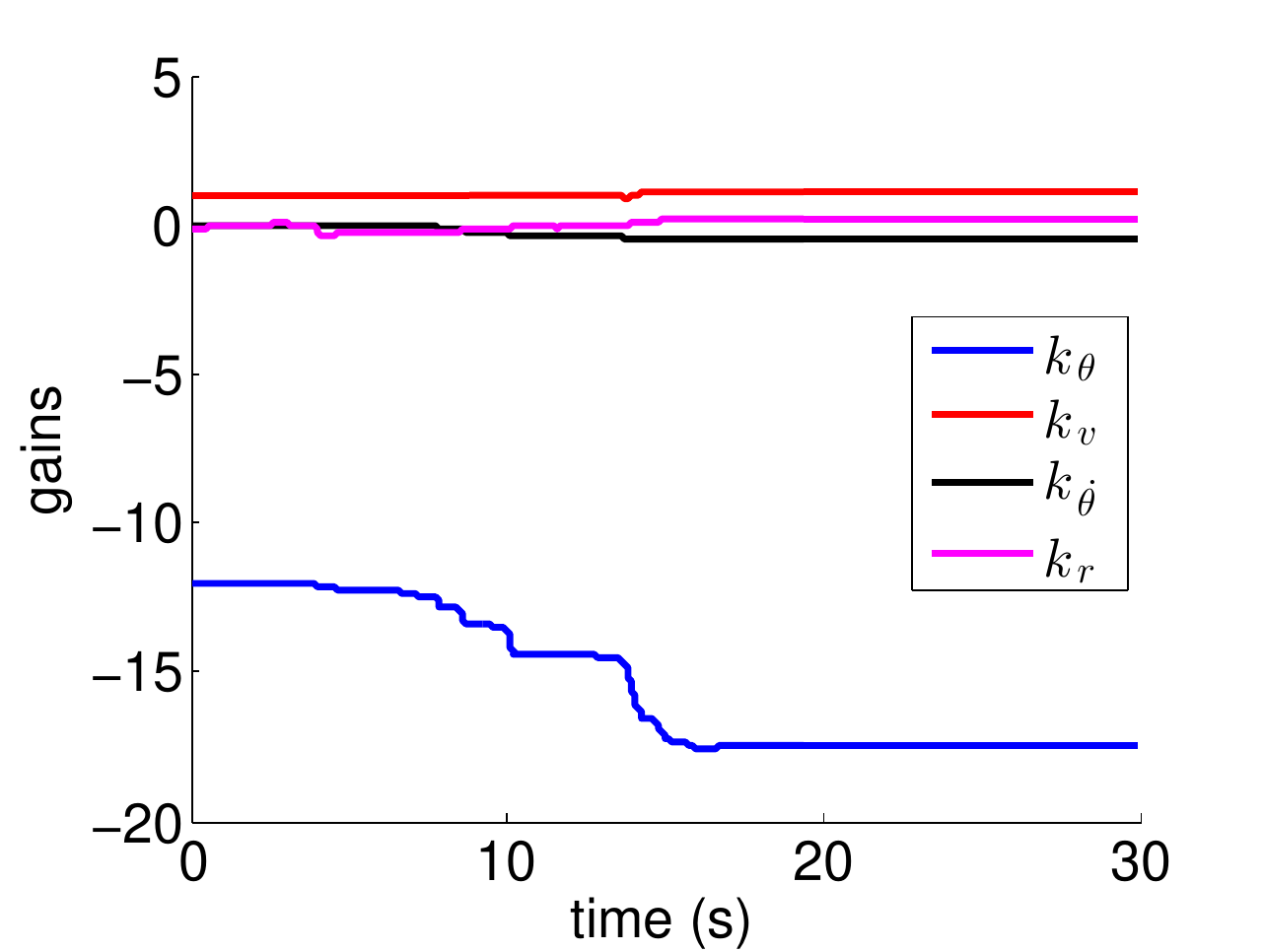} 
\end{minipage}%
\begin{minipage}[h]{0.33\linewidth} 
\centering
\includegraphics[width=0.99\linewidth]{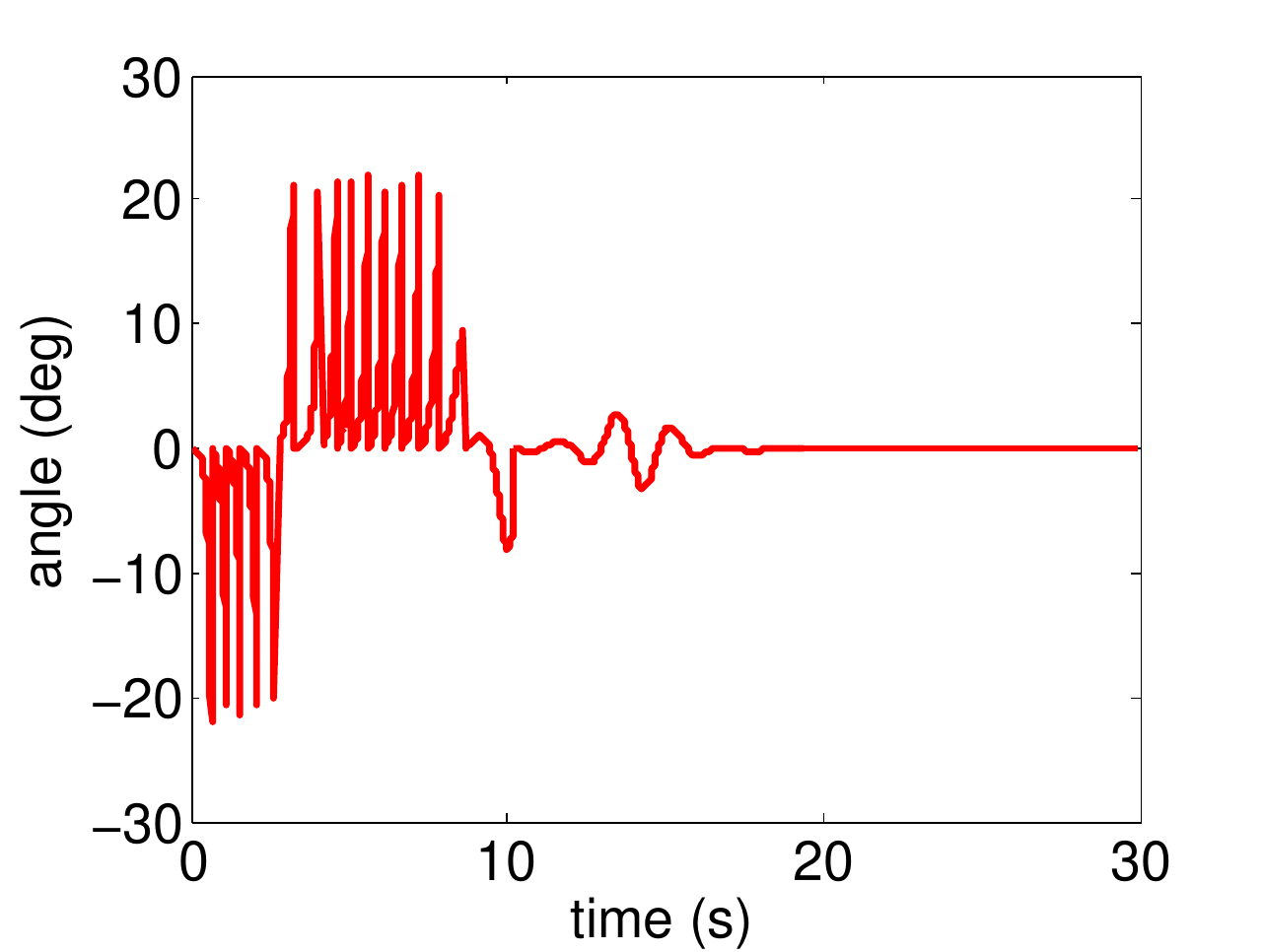} 
\includegraphics[width=0.99\linewidth]{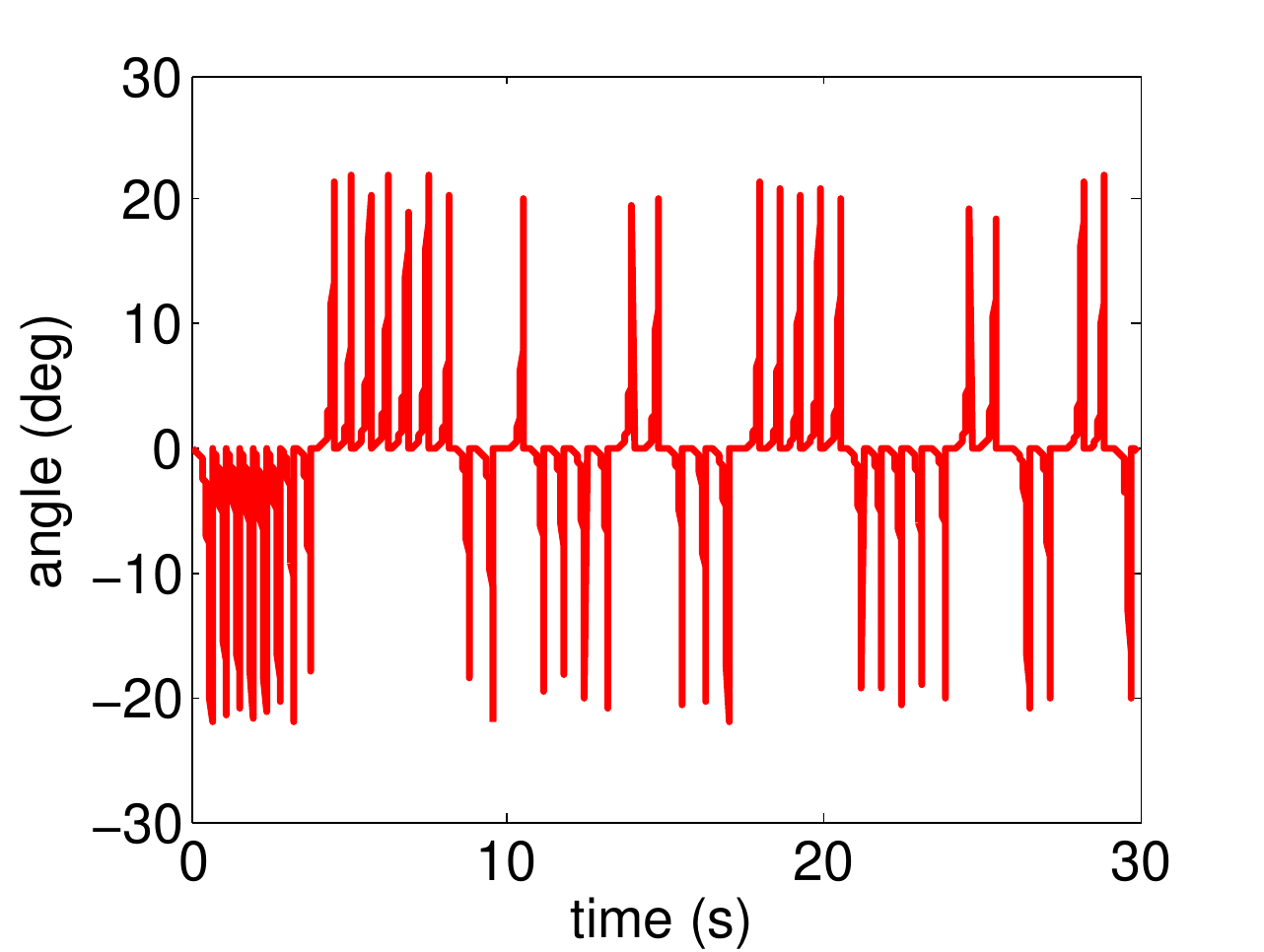} 
\includegraphics[width=0.99\linewidth]{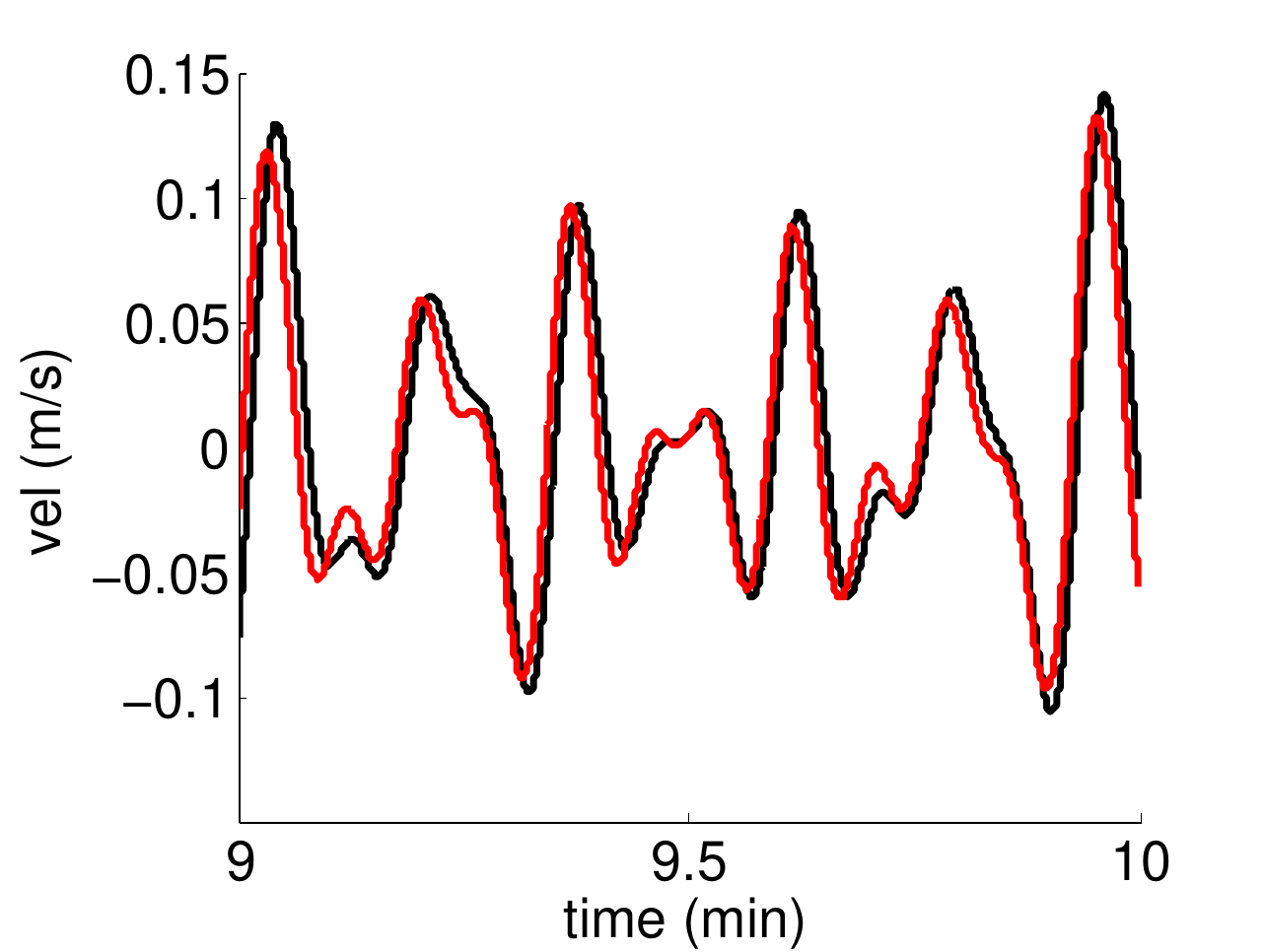} 
\end{minipage}

\caption{ME-LMS applied to a Self-Balancing system. Top row refers to the ME-LMS algorithm and  middle row to the standard LMS. Left panels show the velocity traces (red) and the reference signal (black). Right panels, show the angular position traces. Bottom left: gains of the ME-LMS system during the first 30 seconds of simulation. Bottom right: velocity trace the last minute of 10 mins training of the ME-LMS system.}
\label{fig:FSF-SBS}
\end{figure}

ME-LMS requires approximately 10 seconds for reaching a set of gains that stabilizes the plant following fifteen falls (Fig. \ref{fig:FSF-SBS} \textit{top row}). Standard LMS fails as it converges to a solution that controls for the linear velocity but ignores the angular position (Fig. \ref{fig:FSF-SBS} \textit{middle row}). Indeed, by the end of the 30 seconds of training, standard LMS has reduced the errors in velocity but the speed at which the plant loses balance remains unchanged. Regarding the learning dynamics, we observe how the feedback gains of the ME-LMS change rapidly until the robot maintains balance (Fig. \ref{fig:FSF-SBS} \textit{below left}). After that, the change is gradual but sufficient to achieve following closely the target velocity by the end of the 10 mins training (Fig. \ref{fig:FSF-SBS} \textit{below right}).

\section{Applying ME-LMS to a Bio-Inspired Limb Model}
\begin{figure}
	 \centering
	\includegraphics[width = 0.4\columnwidth]{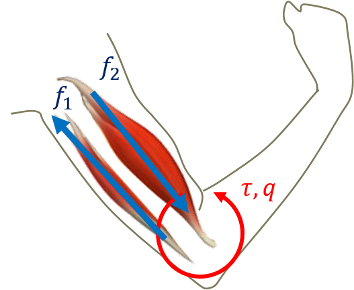}
	\caption{Agonist\--antagonist actuation systems with main variables underlined. $q$ is in both cases the joint angle, and $\tau$ is the external torque. $f_1$ and $f_2$ are the forces exerted by the biceps and triceps respectively.}\label{fig:ago_ant_hum}
\end{figure}

In this section we test the ability of the proposed ME-LMS algorithm to control an antagonistic pair of human muscles acting on a joint, as e.g. the biceps and triceps muscles on elbow joint (Fig. \ref{fig:ago_ant_hum}).

\subsection{Model Derivation}

We consider the muscle model proposed in \cite{gribble1998complex}, as a trade of between accuracy and complexity. According to that, it is possible to describe the force exerted by a muscle as exponential function of its activation $A$, which is considered proportional to the difference between the instantaneous muscle length $l$ and the threshold length $\lambda(t)$, assuming the role of an input. A damping term due to proprioceptive feedback is also considered, proportional to the variation of muscle length $\dot l$. The overall activation is
\begin{equation}\label{eq:activation}
	A(t)=[l(t-d)-\lambda(t)+\mu \, \dot l (t-d)]^+ \, ,
\end{equation}
where $\left[\,x\,\right]^+$ is $0$ when $x\leq 0$, and $x$ otherwise. In the following we neglect the reflex delay $d$. 

Thus considering the forces exerted by biceps $f_1$ and triceps $f_2$ on elbow joint, and the gravity force acting on the forearm, the overall dynamics is 
\begin{equation}\label{eq:dyn}
I \, \ddot{q} + m \, L \, g \, \cos(q) = R(f_1 + f_2),
\end{equation}
where $q$ is the forearm angular position w.r.t.~the arm. $I$ and $m$ are forearm inertia and mass respectively, and $L$ is the distance of forearm center of mass from the joint, all considered constant (i.e. we neglect the dependency from wrist configuration). The two forces exerted by the muscles are 
\begin{equation} \label{eq:forces_m}
	\begin{aligned}
		f_1 &= \, - \rho( e^{\delta A_1}-1), \quad A_1 = [\;\;\; Rq - \lambda_1 + \mu R \dot q ]^+,\\		
		f_2 &= \;\;\; \rho (e^{\delta A_2}-1), \quad
		A_2 = [-Rq + \lambda_2 - \mu R\dot q ]^+ \, , 
	\end{aligned}
\end{equation}
where $\delta$ is a form parameter equal for all muscles, $\rho$ is a magnitude parameter related to force-generating capability. $\lambda_1$ and $\lambda_1$ are the length commands for each muscle. $R$ is the instantaneous lever arm, considered here constant \cite{gribble1998complex}, i.e. $l=Rq$.

At the equilibrium with no external torque \cite{garabini2017soft} the joint angle $q_\mathrm{eq}$, and the stiffness $\sigma$ are
\begin{equation}\label{eq:eq_pos_m}
	q_\mathrm{eq} = \frac{r}{R}, \quad
	\sigma = \left.\frac{\partial \tau}{\partial q}\right|_{q=\frac{\lambda_1 + \lambda_2}{2 \, R}} = 2 \rho \delta R^2 e^{\delta c} \,
\end{equation}
where $r=\frac{\lambda_1+\lambda_2}{2}$ is referred in literature as $r$\--command, and $c=\frac{\lambda_2-\lambda_1}{2}$ is referred as $c$\--command or co\--activation \cite{feldman1995origin}. We consider here $r$ as control input, and $c$ as fixed.

\subsection{Control Derivation}

System \eqref{eq:dyn} in state-space  form is 
\begin{equation}\label{eq:dyn_ss}
	\begin{bmatrix}
		\dot{x}_1 \\ \dot{x}_2
	\end{bmatrix} = 
	\begin{bmatrix}
		x_2 \\ \frac{1}{I} (R (f_1 (x_1,x_2,r) + f_2(x_1,x_2,r)) - m  L  g  \sin(x_1))
	\end{bmatrix}
\end{equation}
where $[x_1, \, x_2] = [q, \, \dot{q}]$. We start by linearizing the system in the equilibrium position $q_\mathrm{eq} = 0$
\begin{equation}
	\begin{bmatrix}
		\dot{{x}}_1 \\ \dot{{x}}_2
	\end{bmatrix} = 
	\begin{bmatrix}
		0 &1\\
		\kappa &\beta
	\end{bmatrix}\begin{bmatrix}
	\dot{{x}}_1 \\ \dot{{x}}_2
	\end{bmatrix} + 
	\begin{bmatrix}
		0 \\ b
	\end{bmatrix}r \, ,
\end{equation}
where
\begin{equation}
	\begin{cases}
		\kappa = -\frac{1}{I}\,(\sigma(c) + m\,L\,g) \\
		\beta = -2 \frac{\rho \mu \delta R^2}{I} e^{\delta c} \\
		b = 2 \frac{\rho \delta R}{I} e^{\delta c}.
	\end{cases}
\end{equation}
Note that $\kappa$ and $\beta$ are negative for any positive choice of the system parameters, and for any value of co-contraction $c$. Thus the system is always locally asymptotically stable in the origin. In the following we consider the problem of trajectory tracking in joint space. Thus the output function is
\begin{equation}
	y = x_1 \, .
\end{equation}

\subsection{Simulations}
The average male forearm weight is $m = 1.36$Kg, and the average male distance of the center of gravity from the elbow joint is $L = 0.155$m \cite{plagenhoef1983anatomical}. The moment of inertia results approximately $I = 0.0109 \, \text{Kg m}^2$. We consider the instantaneous lever arm $R = 1.5$cm. The gravity is approximated to the second digit $9.81 \frac{\text{N m}}{\text{s}^2}$. For the muscle characteristic we consider values in \cite{asatryan1965functional},   \cite{gribble1998complex}: 
$\mu = 0.06$s, $\delta = 0.112 \text{mm}^{-1}$, $\rho = 1$N. Thus
\begin{equation}
\begin{cases}
\kappa = -(4.62 \, e^{0.112 \text{mm}^{-1} c} + 189.72) \frac{1}{\text{s}^2} \\
\beta = -(0.277 \, e^{0.112 \text{mm}^{-1} c}) \frac{1}{\text{s}} \\
b = 308 \, e^{0.112 \text{mm}^{-1} c} \frac{1}{\text{m s}^2}.
\end{cases}
\end{equation}

We consider the following desired swing movement $q_\mathrm{d} = \frac{\pi}{4}\sin(t)$. We consider as reference model the same as before but $10$ times faster, with a response that peaks in 50 ms.

\begin{figure}[h] 
	\centering
	\begin{minipage}[h]{0.33\linewidth} 
		\centering
		\includegraphics[width=0.99\linewidth]{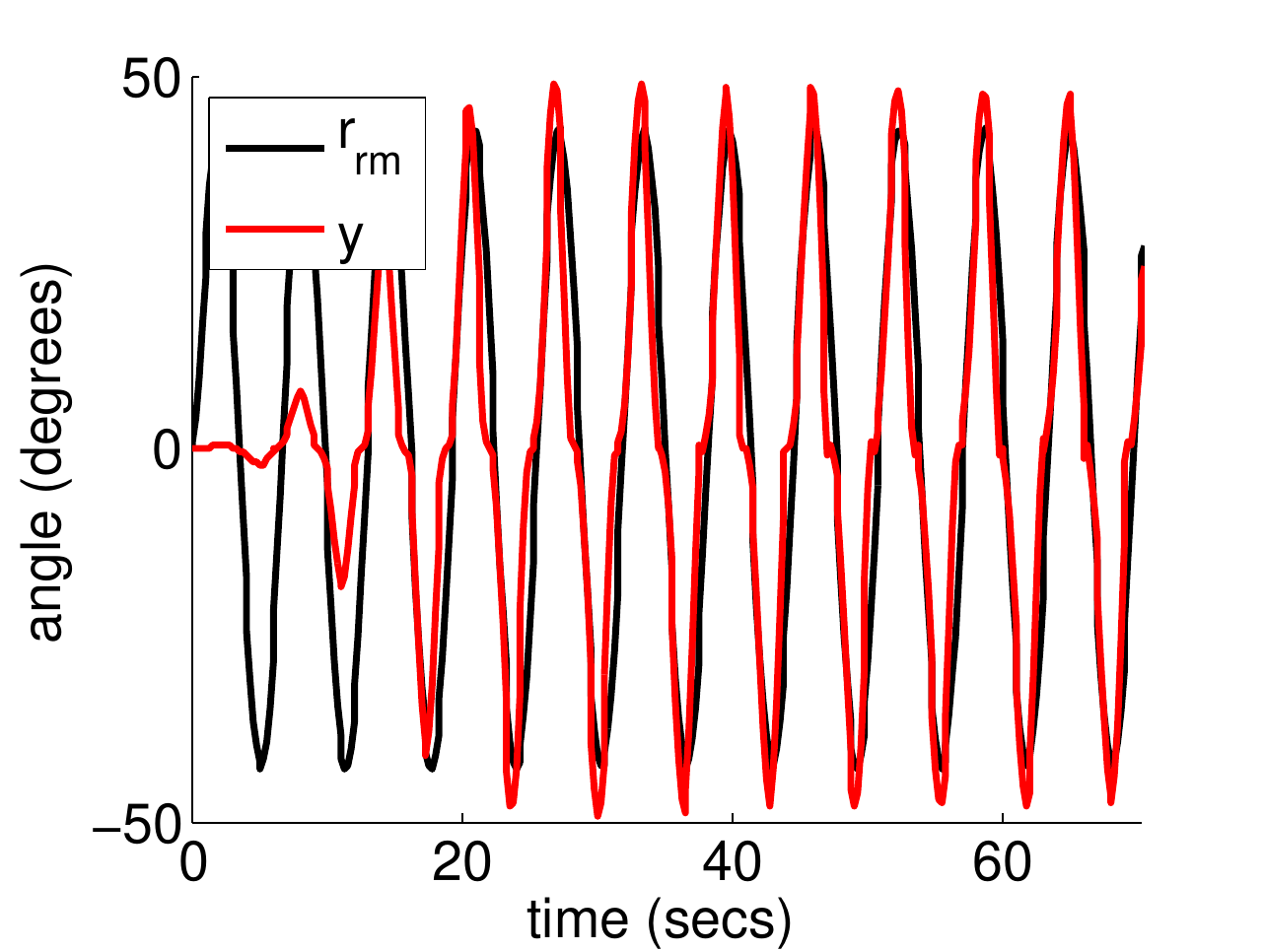}
	\end{minipage}%
	\begin{minipage}[h]{0.33\linewidth} 
		\centering
		\includegraphics[width=0.99\linewidth]{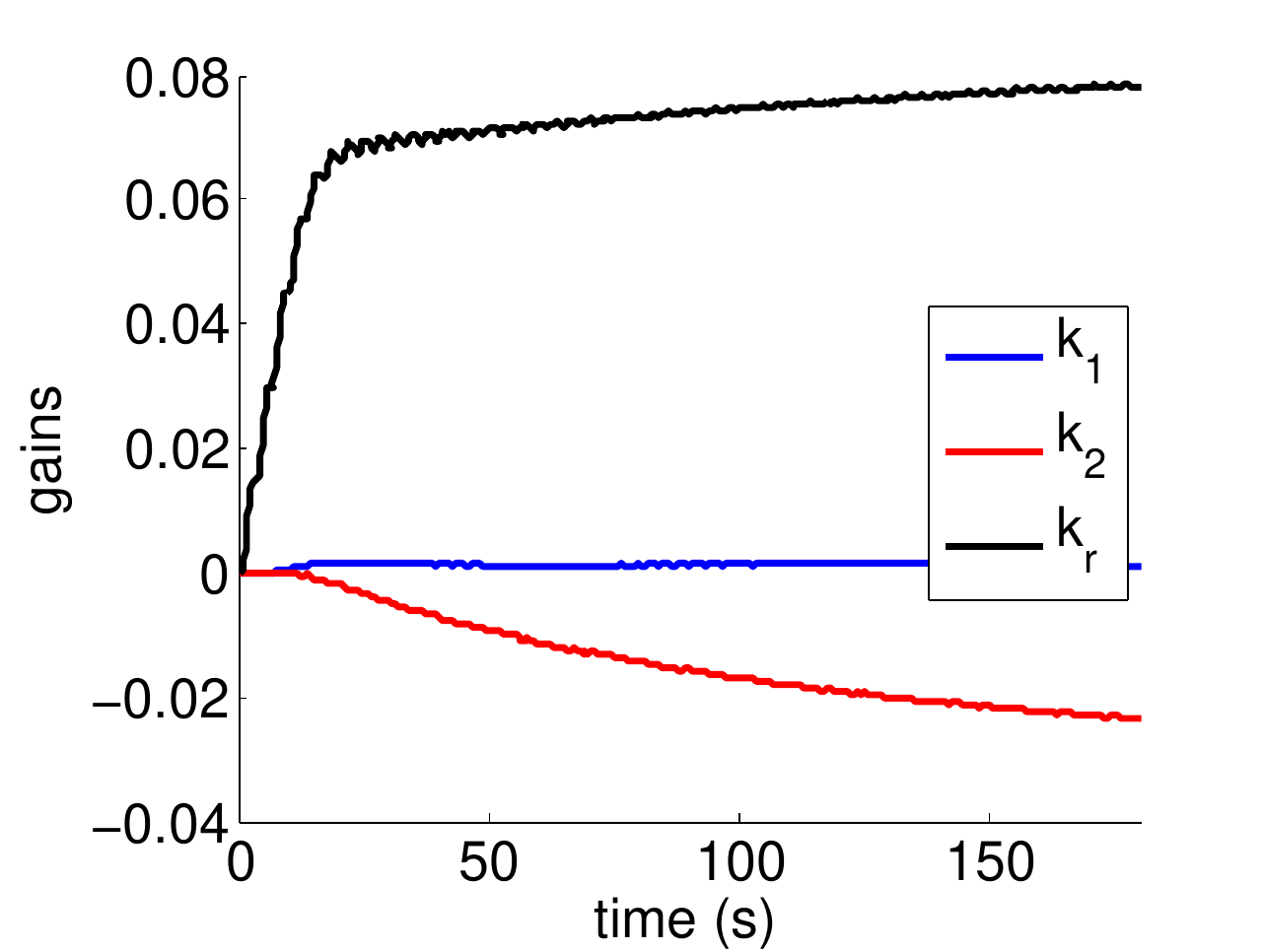}
	\end{minipage}
	\caption{Results of the control of swing movement in a human upper limb, controlled through the proposed cerebellar inspired algorithm. On the left side trajectories are presented, on the right the learning of parameters.} \label{fig:limb_res}
\end{figure}

Note that the system \eqref{eq:dyn_ss} presents many aspects making the problem of controlling it very hard, i.e. it includes exponential and trigonometric actions, the activation terms are not derivable with continuity, and the system can not be written in an affine control form. So the effectiveness of a linear controller in a non-local task is a result not trivial to achieve. We rely here on the natural inspiration of the controller, and on its robust structure already demonstrated in the previous simulations. Fig. \ref{fig:limb_res} presents the simulation results. The algorithm is able to learn the correct control action to track the reference.

\section{Discussion} \label{sec:conclusion}
The cerebellum is a crucial brain structure for accurate motor control. It is phylogenetically old and conserved through evolution in all vertebrates \cite{grillner2011control}. 
Much of the interest that the cerebellum gathered in the machine learning, robotics and adaptive control communities stemmed from its remarkable anatomy \cite{eccles67}, with a general connectivity layout that resembles very closely the one of a supervised learning neural network, such as a Perceptron \cite{rosenblatt1958perceptron}. However, here we have drawn inspiration for a recent insight regarding cerebellar physiology that has emerged simultaneously at both the theoretical \cite{herreros2016forward} and experimental \cite{suvrathan2016timing} domains. That is, that in order to solve appropriately the problem of motor control, neurons from a same type (i.e., the cerebellar Purkinje cells) might display different learning rules in different areas of the cerebellum, matched to the particular behavioral needs \cite{dempsey2016timing}. From a control theory perspective those behavioral needs correspond to the transport latencies and response dynamics associated to the controlled sub-system.

Here we have shown that the model-enhanced least-mean-squares (ME-LMS) learning rule can be easily derived for the task of learning the optimal feedback gains of a fully known plant. Second, we have shown  that the ME-LMS learning rule converges in a series of tasks in which conventional LMS would fail, as is the case of a non-minimum phase plant. 
 
Regarding the derivation of ME-LMS presented here, it is worth noting that although a similar result was originally obtained in \cite{herreros2016forward} using a cerebellar-based control architecture, the derivation presented here applies to two very general control architectures; namely, proportional full-state-feedback and proportional error-feedback control. Hence, in that sense, the current derivation is cerebellar-independent.


\section{Conclusions and Future Work}

In this work we proposed a control algorithm in which a linear feedback action is combined with an adaptation rule for the control gains inspired from cerebellar learning in animals. 
The controller is also analytically derived as gradient descent estimation of feedback gains for LTI systems.

We tested the effectiveness of the algorithm in three different simulation scenarios, including the control of a model of human upper limb, closing the loop with the biological inspiration. The algorithm presented better performance w.r.t. the classic LMS rule.
In future work we plan to experimentally test it on bio-inspired robotic systems.

Although the algorithm presented in practice a stable behavior, to analytically prove the stability of the closed loop system is a challenging task. This is due not only to the non-linearities introduced by the possibility of adapting control gains (common in the context of adaptive control), but also to the strong interplay between the three dynamics involved: the system, the eligibility trace, and the control gains. However we consider this step very important for the full formalization of the proposed learning rule, and so we depute this study to future works.

\section*{Acknowledgments}
The research leading to these results has been supported by the European Commission's Horizon 2020 socSMC project (socSMC-641321H2020-FETPROACT-2014), the European Research Council's CDAC project (ERC-2013-ADG 341196),  the European Commission's  Grant no. H2020-ICT-645599 SOMA: SOft MAnipulation and the ERC Advanced Grant no. 291166 SoftHands.


\bibliographystyle{plain}
\bibliography{biblio}

\end{document}